# Discovery and effective application of magnetotelluric in the exploration of geothermal resources in Banqiao depression

REN XiaoQing[a,b]   WANG HongLiang[a]   GAO XiaoRong[b]   ZHAO Xin[b]

[a] School of Energy Resources, China University of Geosciences (Beijing), Beijing 100083, China
[b] Sinopec Green Energy Geothermal Development Co., Ltd., Baoding 071800, China

*Corresponding author email:renxiaoqing.xxsy@sinopec.com

**Abstract**: This article aims to solve the local geothermal development problem by conducting geothermal exploration in Xiaowangzhuang Town, Banqiao Depression, Bohai Bay Basin. The pore type thermal reservoir in this area is mainly composed of the sandstone of the Neogene Guantao Formation. By using the magnetotelluric method (MT) and starting from structural analysis in the deep part, a total of 2 MT survey lines were laid out, with a total length of 9.0km. After collecting, processing, inverting, and interpreting field data, a total of three fault structures were discovered based on the inversion interpretation results, which are basically consistent with regional structures. The main controlling structure is the Cangdong Fault. The burial depth of thermal reservoirs in the region has been determined. On this basis, the design and construction of one geothermal well were carried out, with a designed vertical depth of 1450m and a measured depth of 1630m. The water temperature at the wellhead of the geothermal well was designed to be≥55±3°C. When the depth reduction is ≤50m, the water output from the wellhead is≥80±5m³/h. The actual drilling depth was 1445.6m (vertical depth)/1630.00m (depth measurement), with a drop of 36.8m. The water volume was 89m³/h and the water temperature was 51°C, which effectively verified the practicality and effectiveness of MT in detecting geothermal water in the Banqiao Depression of the Bohai Bay Basin.

# 1  Introduction

In recent years, in order to pursue sustainable low-carbon emission energy, the global exploration and development of geothermal energy has continued to increase (Lund et al., 2005; Wang Shejiao et al., 2024). Tianjin is located in the northeast of Bohai Bay Basin. After 50 years of geothermal exploration and development, 10 geothermal anomaly areas have been delineated and 8 geothermal fields have been explored (Ru Hongjiu et al., 2018; Tang Yongxiang et al., 2024; Tang Yongxiang et al., 2023). Banqiao depression is located in the southwest of Tianjin, which has thick clastic sedimentary strata and high geothermal flow, but Banqiao depression has always been a blank area for geothermal exploration and development.

In order to change the energy structure of winter heating in Xiaowangzhuang town in Banqiao depression from natural gas as the main heat source to geothermal heat source, it is planned to explore geothermal resources and drill geothermal wells in Xiaowangzhuang town in Banqiao depression. In the whole preliminary exploration process, we need to make clear which strata will be drilled in the planned drilling place and identify the main thermal reservoirs, which provide the foundation and basis for the design of geothermal wells. Therefore, in this study, the following problems are mainly explored: (1) the use of geophysical methods and how to arrange geophysical survey lines, and how to collect and analyze data. (2) What strata are included in the stratigraphic structure of this area? (3) Are there any major faults in the study area? These are the geophysical exploration workflow and main influencing factors for evaluating the thermal storage resources of porous sandstone in sedimentary basins.

Magnetotelluric sounding method is a geophysical exploration method to study the electrical structure of the earth by using natural alternating electromagnetic field (Rosenkjaer et al., 2015; Constant width et al. Sun Huanquan et al., 2024; Wu Jiawen et al., 2023), by studying the electromagnetic data collected on the surface, the information of resistivity distribution of underground media at different depths can be deduced (Chave et al., 2012; Rosenkjaer et al., 2015; Shah et al., 2015)。 In the process of geothermal development in Tianjin, researchers mainly studied the tectonic framework and faults (Sui Shaoqiang et al., 2019) and thermal reservoir evaluation. Wumishan Formation of Jixian system is widely distributed in Tianjin, and the Cambrian and Ordovician thermal reservoirs are largely absent in the core of Cangxian uplift, and widely distributed in other areas, which is one of the main thermal reservoirs (Li Shengtao et al., 2022; Tang Yongxiang et al., 2023; Wu Jiawen et al., 2023; Yao Yahui et al., 2022; Yue Dongdong et al., 2023; Yue Dongdong et al., 2020), the Neogene Guantao Formation is mainly concentrated in wuqing district and Binhai New Area (Li Shan et al.,2023; Ruan Chuanxia et al.,2018,2023; Shen Jian et al.,2005).

In order to find out the distribution of Guantao Formation sandstone thermal reservoir in Banqiao Depression, two survey lines were laid in Xinyuanli and Xiangyangli communities of Xiaowangzhuang Town in Banqiao Depression, and the data were collected, processed, inverted and interpreted, so as to find out the buried depth of the thermal reservoir in the region and identify the faults in the region. Based on this, the design and construction of geothermal wells were carried out, which achieved ideal results. The above practice proves the effectiveness of this method in identifying the middle and deep geothermal resources.

## 2 Geological and geothermal settings

### 2.1 Areal geology

The study area is located in Banqiao Depression , a class I structural unit, a class II structural unit in the northern margin of North China Platform, a class III structural unit in the North China Fault Depression and a class IV structural unit in Huanghua Depression. The working area is located in the southwest of Banqiao depression, with Cangdong fault as the boundary and Cangxian uplift in the northwest (Cheng Wanqing et al., 2012). Affected by the continuous activity of the northwest boundary fault, the whole fault depression is a dustpan-shaped fault depression with NNE distribution, which is steep on the west side and gradually flattens to the east. The main control fault developed in the working area is Cangdong fault, which is located in the west of the working area, and the overall strike is NNE and SEE, and it is a normal fault with relatively rising west plate (Song Kun et al., 2016). It is the boundary fault between Cangxian Uplift and Huanghua Depression, and the structural plane of the fault changes greatly, and the cross section is mostly shovel-shaped.

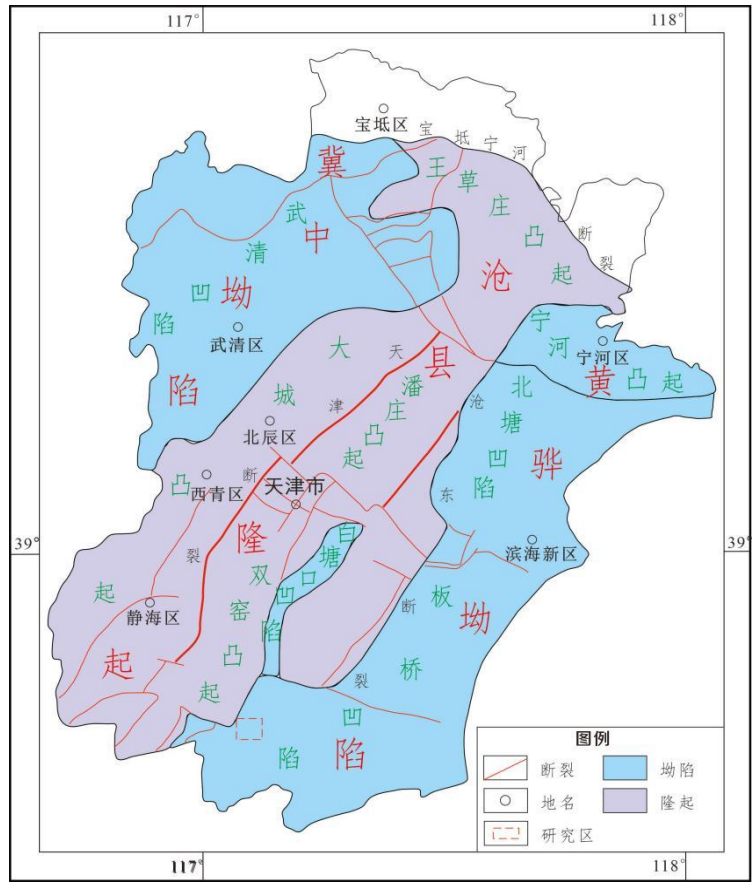

Fig. 1 Regional Geological Map of the Study Area

## 2.2 Stratigraphic characteristics

The Paleogene Shahejie Formation, Dongying Formation, Neogene Guantao Formation and Minghuazhen Formation are mainly developed in Banqiao Depression, which constitute huge clastic rock deposits and provide favorable reservoir space for geothermal water. Cangxian uplift is on the west side of Cangdong fault, and the bedrock is dominated by Paleozoic, and the Paleozoic uplift on the west side is obvious. Huanghua depression is in the east, and the bedrock is Mesozoic, and the Mesozoic tends to thicken eastward (Li Shengtao et al., 2022). Cangdong fault mainly controls the Mesozoic and Cenozoic deposits in this area, and the accumulated Paleogene Kongdian Formation is very thin, mainly Shahejie Formation and Dongying Formation, with a thickness of about 3000～5000m m. The strata in the study area developed from old to new in turn, including Mesoproterozoic, Paleozoic, Mesozoic and Cenozoic (Meng et al., Korea, 2020; Liu Jinxia & Jin Zhijun, 2004; Shi Qianru et al.,2020; Zhao Na et al.,2016).

## 2.3 Temperature field characteristics

Controlled by deep crustal structure, basement structure, some large faults and magmatic activity, the geothermal field in Huanghua Depression is obviously divided into three parts in the plane (Zhang Baiming, 2007). In the middle of Huanghua Depression, Banqiao Depression and other places, the geothermal gradient belongs to the high temperature area, and it is low in the west and high in the east, high in the south and low in the north. According to the geothermal data of the caprock in Xiaowangzhuang Town and its nearby area, the geothermal gradient in the study area is between 2.5-4.0℃/100m (Ruan Chuanxia, 2018). It belongs to the middle temperature zone in the caprock of Tianjin area.

## 3 Geophysical working method

### 3.1 Method principle

Magnetotelluric sounding is a geophysical exploration method that uses natural alternating electromagnetic field to study the earth's electrical structure. It has the characteristics of low frequency, long wavelength, convenient operation, no high resistance shielding and large detection depth, and is mainly used to detect the underlying structure in the high resistance layer coverage area (Cao Xiaoling et al.,2017; Li Bin et al.,2010; Chang Kuan et al.,2018; Sun Huanquan et al.,2014; Wu Jiawen et al.,2023; Ruan Chuanxia, 2018), local structural morphology, fault properties and distribution characteristics, investigation of geothermal fields, etc. (Cao Xuegang et al.,2021; Deng Yan et al.,2024; Lei Qing et al.,2024; Zhao Jianliang et al.,2010).

### 3.2 Data acquisition

This time, two MT survey lines were designed, with 39 measuring points with directions of 135, with a distance of 300m, and the cumulative length of two sections was 9.0km. In order to obtain the development of deeper underground structures and strata, according to some achievements of magnetotelluric sounding, some measuring points were observed for a long time, and the observation time was as long as 8 hours.

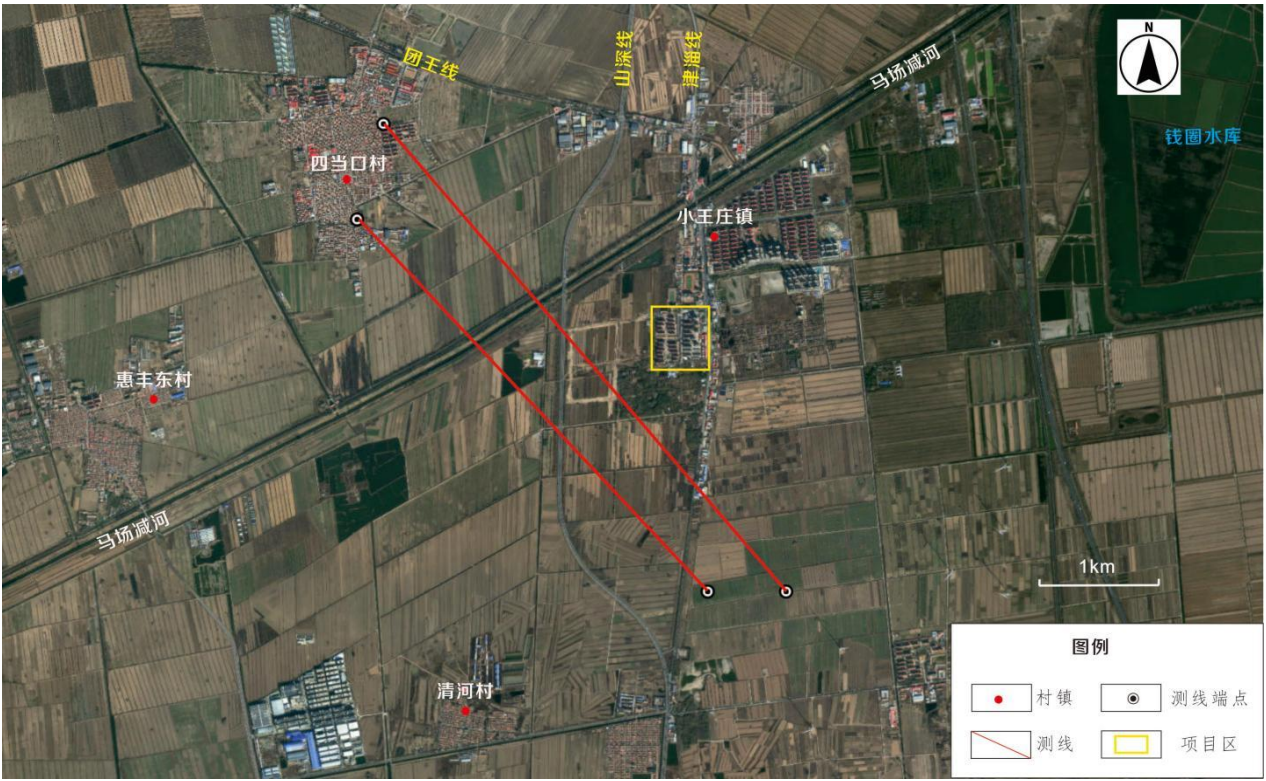

Fig.2 Layout of survey lines in the stdudy area

## 3.3 Data processing

The magnetotelluric signals observed in the field can only be interpreted after a series of indoor processing. It can be said that data processing is the most important link of magnetotelluric sounding method. Data processing generally follows several principles: the principle from known to unknown; The principle from qualitative to quantitative; From coarse to fine, the principle of gradual deepening, etc.

### 3.3.1 Impedance tensor

The purpose of decomposition of magnetotelluric impedance tensor is to eliminate the influence of local distortion from the observed magnetotelluric response and obtain parameters such as regional tectonic impedance and strike. This method is widely used in magnetotelluric sounding data processing (Cao Xuegang et al., 2021; Deng Yan et al., 2024; Lei Qing et al., 2024; Zhao Jianliang et al., 2010). Whether the two-dimensional magnetotelluric inversion is effective depends on the tensor impedance decomposition results, and the reasonable tensor impedance decomposition results make the TE and TM polarization modes have practical physical significance. In order to carry out two-dimensional inversion effectively, the validity of Swift tensor impedance decomposition method is applied.

Swift decomposition method, also known as traditional decomposition method (Chen Xiaobin, 2005), rotates the measured XY plane clockwise around the Z axis, so that the diagonal element of the impedance tensor is zero. Considering that the actual data generally contains errors, it is impossible to accurately equal zero. Generally, the diagonal element of the impedance tensor is minimized or the anti-diagonal element is maximized to obtain the impedance of the principal axis orientation (Chen Xiaobin et al., 2021).

$$|Z'_{XX}|^2 + |Z'_{YY}|^2 \to \min \text{ 或者 } |Z'_{XY}|^2 + |Z'_{YX}|^2 \to \max \tag{1}$$

When equation (2) is satisfied, it indicates that the impedance has rotated to the principal axis direction of the electrical structure, and θ0 at this time is the principal axis azimuth of the regional two-dimensional structure, and its expression is (Swift, 1967; Cai Juntao et al., 2010; Yin Yaotian et al., 2012):

$$\theta_0 = \frac{1}{4}\arctan\frac{(Z_{XX}-Z_{YY})(Z_{XY}+Z_{YX})^* + (Z_{XX}-Z_{YY})^*(Z_{XY}+Z_{YY})}{|Z_{XX}-Z_{YY}|^2 - |Z_{XY}+Z_{YX}|^2} \tag{2}$$

The angle is 90 fuzzy, so it cannot be determined whether it is strike angle or dip angle. When θ0 satisfies the following formula.

$$\tan 4\theta < \frac{(Z_{XY}+Z_{YX})^*(Z_{XY}+Z_{YX}) - (Z_{XX}-Z_{YY})^*(Z_{XX}-Z_{YY})}{(Z_{XY}+Z_{YX})^*(Z_{XX}-Z_{YY}) + (Z_{XX}-Z_{YY})^*(Z_{XY}+Z_{YX})} \tag{3}$$

The angle sought is the angle when, $|Z'_{XY}|^2 + |Z'_{YX}|^2 \to \max$ which is the principal axis angle; Otherwise θ₀ 90 is the principal axis angle. So the impedance tensor in the direction of the main axis of the structure is.

$$Z = \begin{bmatrix} Z_{XX} & Z_{XY} \\ Z_{YX} & Z_{YY} \end{bmatrix} \tag{4}$$

The spindle azimuth θ₀ obtained by Swift rotation method is directly obtained from the observed impedance tensor z, and when calculating θ₀, only the observation data and rotation angle in the observation coordinate system are involved, without any additional constraints. When the actual structure is not two-dimensional, the impedance tensor matrix is still a full tensor matrix in the rotated coordinate system, and the principal axis angle obtained at this time is actually a comprehensive principal axis angle closest to the two-dimensional structure. If the actual electrical structure is a standard two-dimensional structure and there is no local distortion, the impedance tensor matrix after rotation is an anti-angle matrix, and θ₀ obtained at this time is the principal axis azimuth of the regional structure. Because this method involves the structural principal axis, it has the best adaptability to the structural model with clear strike or tendency.

The decomposition result of Swift 2D deviation in the exploration area is shown in Figure 1, which shows the distribution of 2D deviation values of all measuring points in the exploration area. When it is greater than 0.01Hz, Swift 2D deviation is almost less than 0.4, indicating that the geological structure of the exploration area as a whole is biased towards 2D.

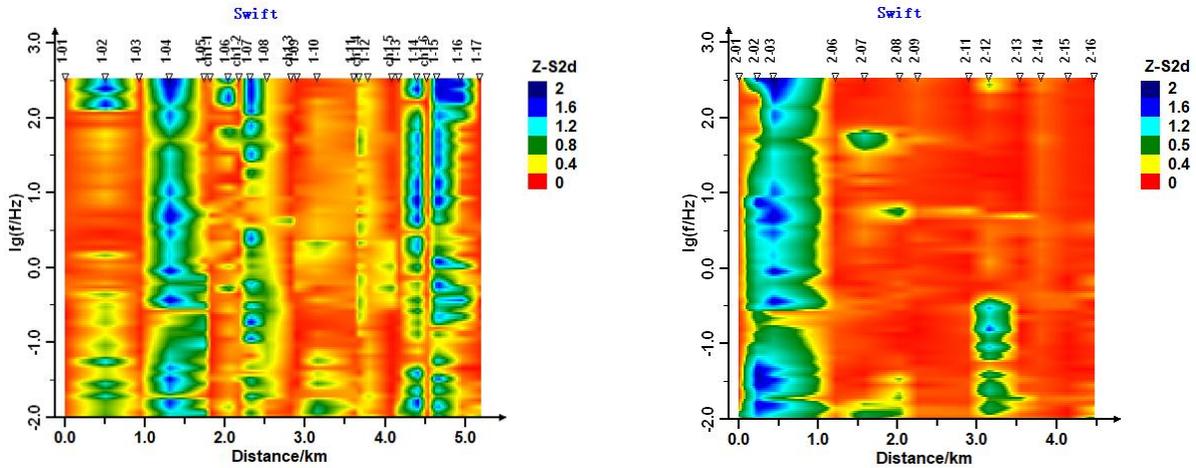

Fig. 3 Proposed Cross Section of Swift 2D Deviation in the Exploration Area
(a) L1 measurement line Swift two-dimensional deviation profile (b) L2 measurement line Swift two-dimensional deviation profile

### 3.3.2 Inversion method

At present, the common one-dimensional inversion methods include one-dimensional adaptive regularization inversion method, Bostick method, RouPlus method and OCCAM method (Bostick,1977; Constable et al.,1987; Chen Xiaobin et al.,2005; Tan Jie, 2015; Xu Shizhe, 1995; Xu Yucong et al.,2015., Yang Yuanhong., 2020; Zhang Juntao et al.,2015), this time, the adaptive regularization inversion method is adopted, which can better reflect the low-resistivity anomaly bands and has a certain vertical stratification trend.

This time, by means of magnetotelluric sounding inversion, the data in the work area are calculated with different dimensions, different modes and different parameters, and a large number of inversion models are compared and analyzed, and finally a reliable underground electrical structure model is determined. There are many two-dimensional inversion methods, among which fast relaxation inversion method and nonlinear conjugate gradient inversion method are representative (Hu Jiashan et al.,2009; Zhang Kun et al.,2013). In this inversion, the nonlinear conjugate gradient (NLCG) two-dimensional magnetotelluric sounding inversion algorithm is used to inverse the sounding data with different modes (such as TM, TE, TM+TE, etc.) and different parameters, and a large number of two-dimensional electrical structure models along the profile are obtained.

Taking L2 as an example, the two-dimensional inversions of TM, TE and TM+TE are made respectively, as shown in the following figure. Through comparison, it is found that the two-dimensional inversion map in TM mode can show certain electrical stratification characteristics, but the low resistance anomaly near the fault structure is not obvious. TE mode can clearly show the low-resistance anomaly, but the overall resistivity is low, and the low-resistance anomaly in the middle of the profile from 1000m to 2000m is too enlarged. TM+TE mode can well reflect the characteristics of electrical stratification, and also can well show low-resistance anomalies, so L2 finally chose the two-dimensional inversion map of TM+TE mode as the result map.

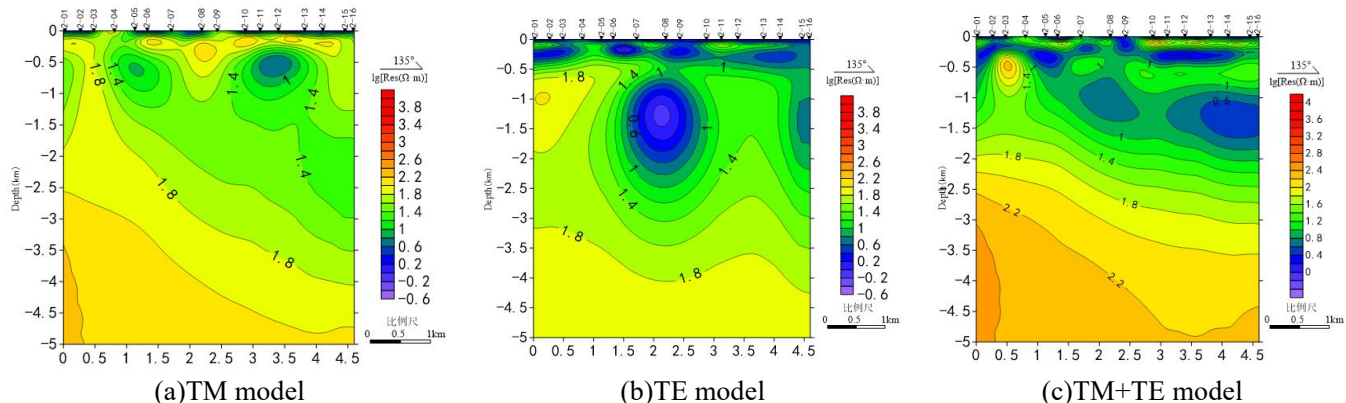

(a)TM model　　　　　　　　　(b)TE model　　　　　　　　　(c)TM+TE model
Fig. 4 Two dimensional inversion maps under different modes

## 4 Results and discussion

## 4.1 Comprehensive structural interpretation

（1）L1

According to the frequency-apparent resistivity diagram of L1 line (Fig. 3), there are two obvious low-resistance bands between point 4 and point 8, and correspondingly, there are obvious abnormal reactions at the same position in the impedance phase diagram. According to the one-dimensional inversion profile, there are many low-resistivity traps between No.4 and No.13 points, which makes the apparent resistivity in this area obviously lower than that in the surrounding area. According to the two-dimensional inversion map, there is an obvious low stop band between points 4 ~13, accompanied by the undercut of isoline. From the above-mentioned maps, it can be seen that there is a large-scale fault from No.4 to No.13, which inclines to the large point and is named F1. The overall dip angle is estimated to be about 75. Combined with the analysis of geological data, this fault should be Cangdong fault.

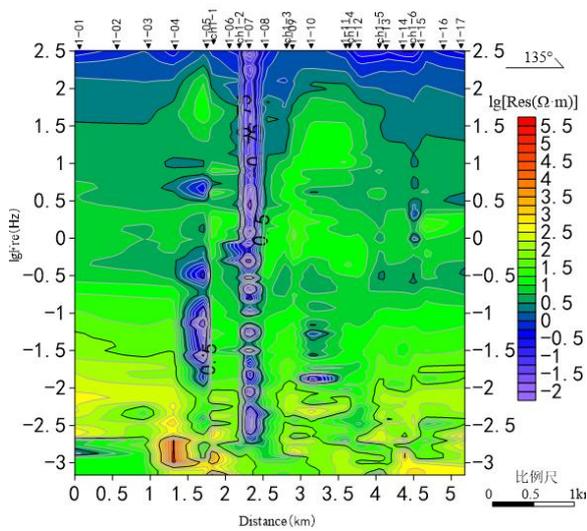 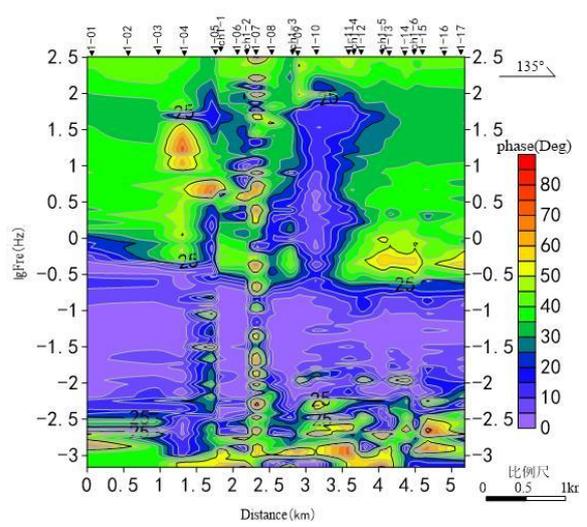

(a) Frequency apparent resistivity contour map　　(b) Frequency impedance phase contour map

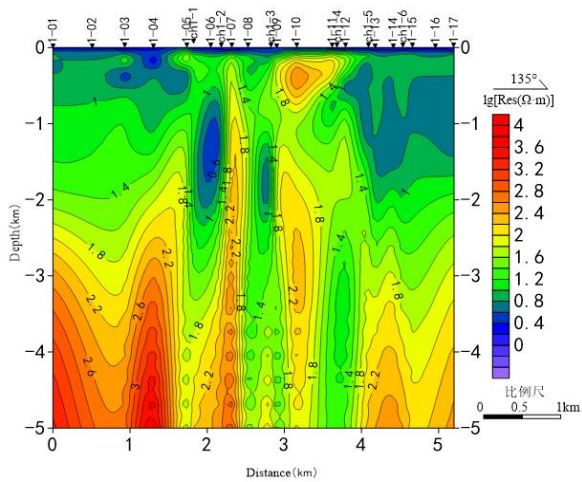 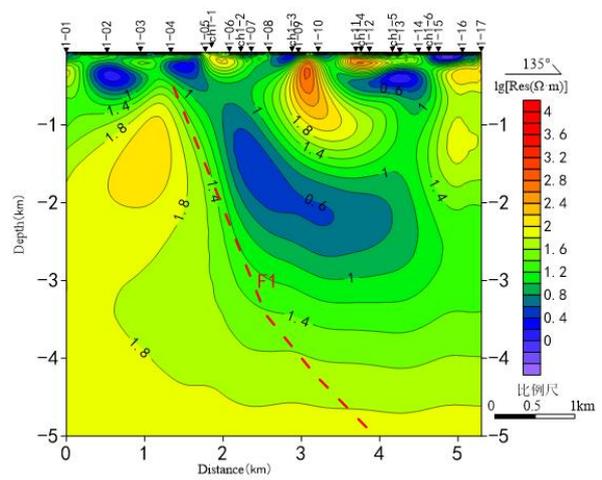

(c) One-dimensional inversion graph　　(d) 2D inversion map

Fig.5 Comprehensive Results of L1 MT

（2） L2

According to the frequency-apparent resistivity diagram of L2 survey line (Fig.4), there are obvious low-resistance anomalies near Point 7 and Point 12, and correspondingly there are abnormal reactions at the same position in the impedance phase diagram. According to the one-dimensional inversion profile, there are many low-resistance areas and isoline undercuts between No.4 and No.12 points, and many isoline undercuts in this area tend to deepen from small to large points. According to the two-dimensional inversion map, there is also an obvious overall contour undercut in this area. From the above-mentioned various maps, it can be seen that there is a large-scale fault at Point 4, which is inclined to the large-scale point, that is, F1 (Cangdong fault), which strikes northeast and tends to southeast.

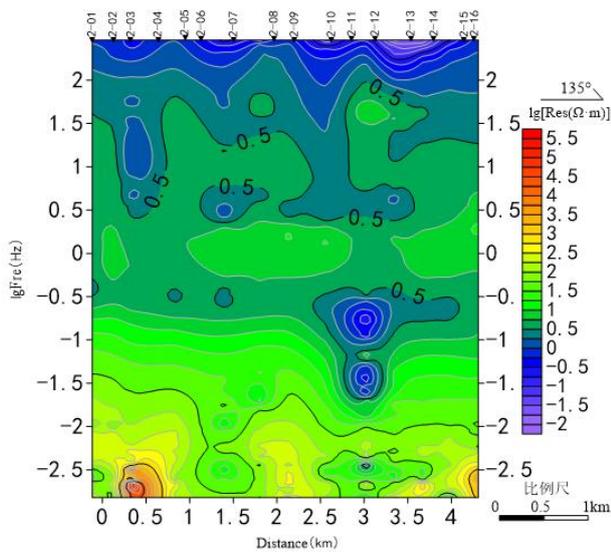
(a) Frequency apparent resistivity contour map

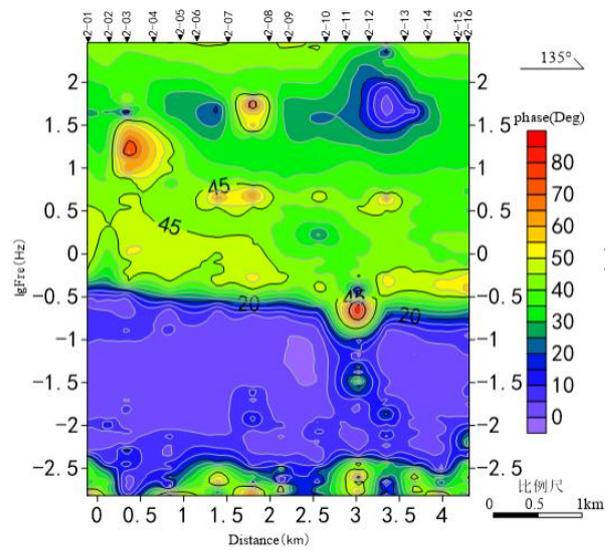
(b) Frequency impedance phase contour map

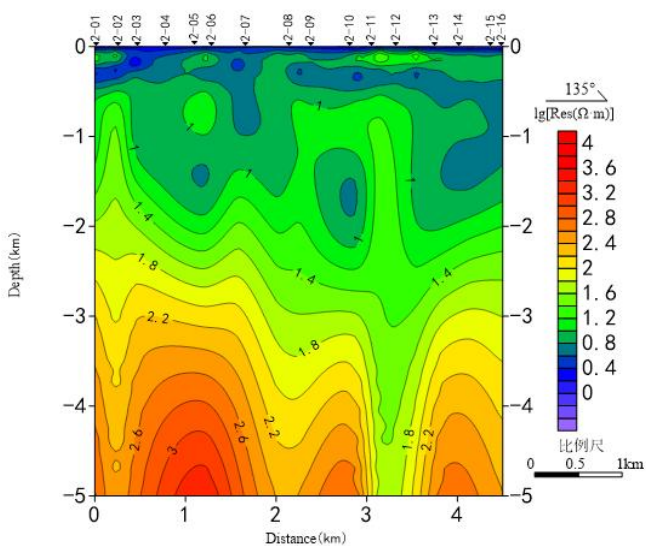
(c) One-dimensional inversion graph

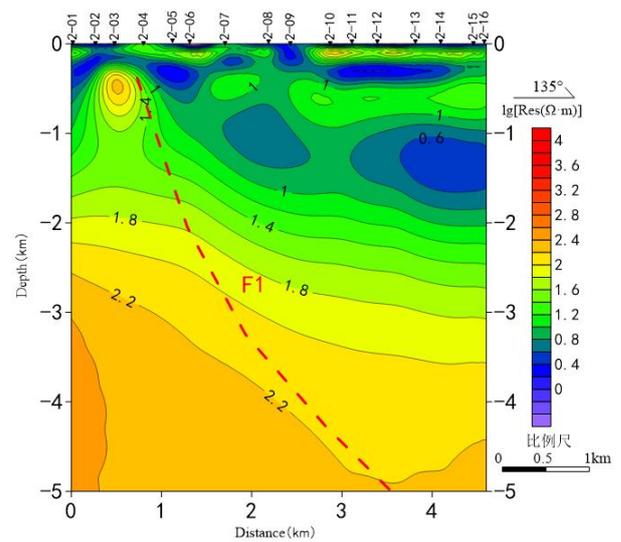
(d) 2D inversion map

Fig. 6 Comprehensive Results of L2 Magnetotelluric Sounding

## 4.2 Geological interpretation

Influenced by Cangdong fault, the section of L1 2D inversion map was cut by the fault, and the overall resistivity at 3000m was low, and the overall horizontal stratification was not obvious, but the resistivity on both sides of the fault was obviously demarcated. This L1 profile stratigraphic interpretation has divided Guantao Formation floor and Paleogene floor. The buried depth of Guantao Formation floor is more than 1000m, and that of Paleogene floor is more than 3000m m.. With Cangdong fault as the boundary, Paleozoic and Proterozoic are on the left, and Paleogene and Mesozoic are on the right, which is consistent with the reflection on the two-dimensional inversion map, and the strata on both sides of the fault are different (Fig.7). L1 stratum is divided into: the buried depth of Quaternary floor is 400~500m. The buried depth of Neogene floor is 1400~1500m east of Cangdong fault and 1300~1400m west of Cangdong fault. Paleogene is distributed in the east of Cangdong fault, and the deepest floor is about 3600m m. The buried depth of Wumishan Formation in Jixian system is more than 5000m east of Cangdong fault.

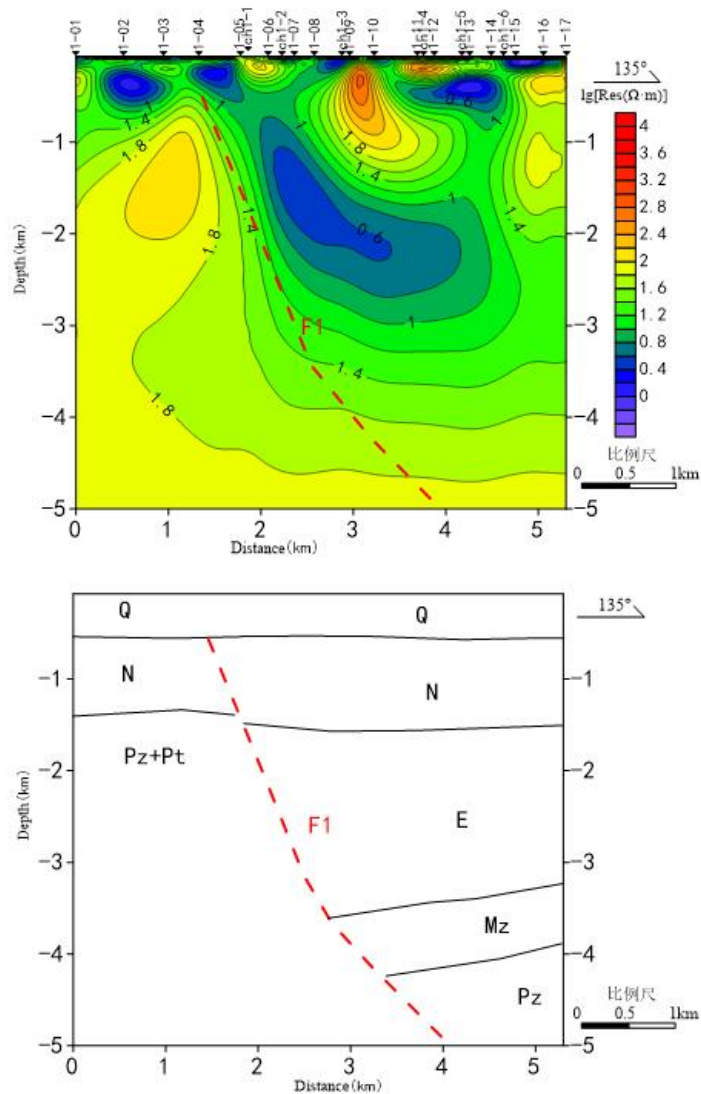

Fig. 7 Interpretation of L1 Formation Profile

L2 is bounded by No.4 point, and the overall apparent resistivity on the left side is greater than that on the right side, which is clearly reflected in the two-dimensional inversion diagram, indicating that the strata development on both sides of the fault are different. L2 profile strata are divided into Guantao Formation floor and Paleogene floor. The buried depth of Guantao Formation floor is more than 1000m, and that of Paleogene floor is more than 3000m m.. With Cangdong fault as the boundary, Paleozoic and Proterozoic are on the left, and Paleogene and Mesozoic are on the right, which is consistent with the reflection on the two-dimensional inversion map, and the strata on both sides of the fault are different (Fig.8). L2 stratum is divided into: the buried depth of Quaternary floor is 400~500m. The buried depth of Neogene floor is 1400~1500m east of Cangdong fault and 1300~1400m west of Cangdong fault. Paleogene is distributed in the east of Cangdong fault, and the deepest floor is about 3600m m. The buried depth of Wumishan Formation in Jixian system is more than 5000m east of Cangdong fault.

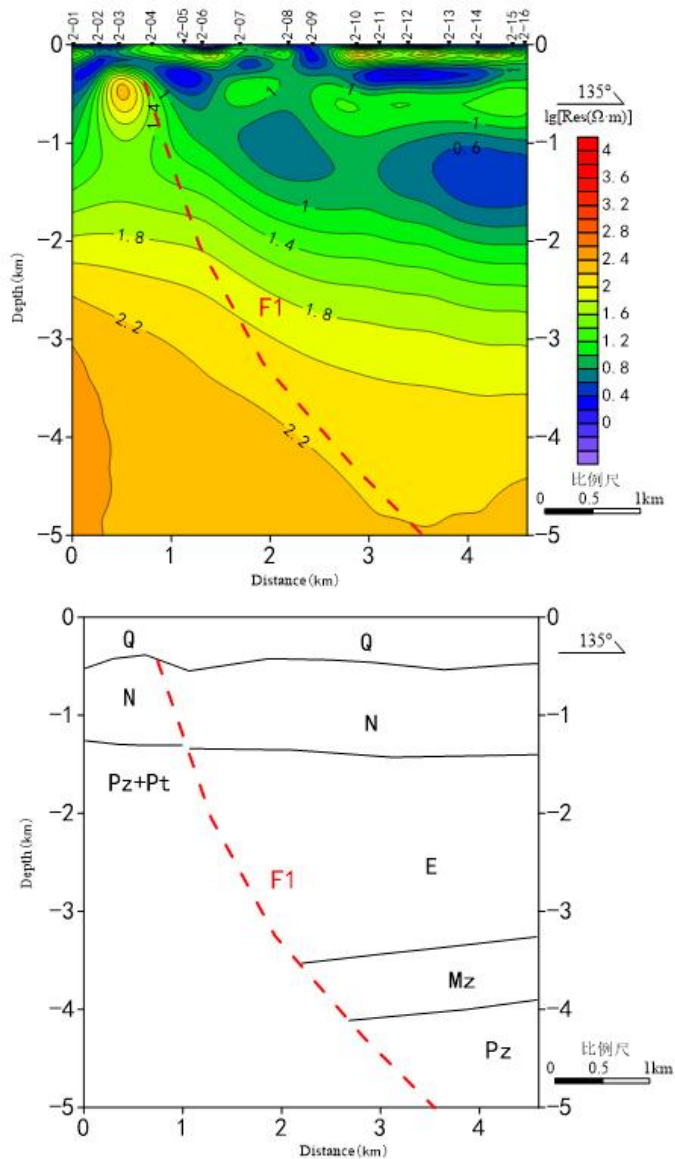

Fig.8 Explanation of L2 stratigraphic profile

Based on the results of this magnetotelluric sounding, combined with the comprehensive analysis of seismic interpretation results, surrounding borehole data and geological data, the AA' geological profile map (Fig.9) and the buried depth contour map of Guantao Formation floor (Fig.10) are made.

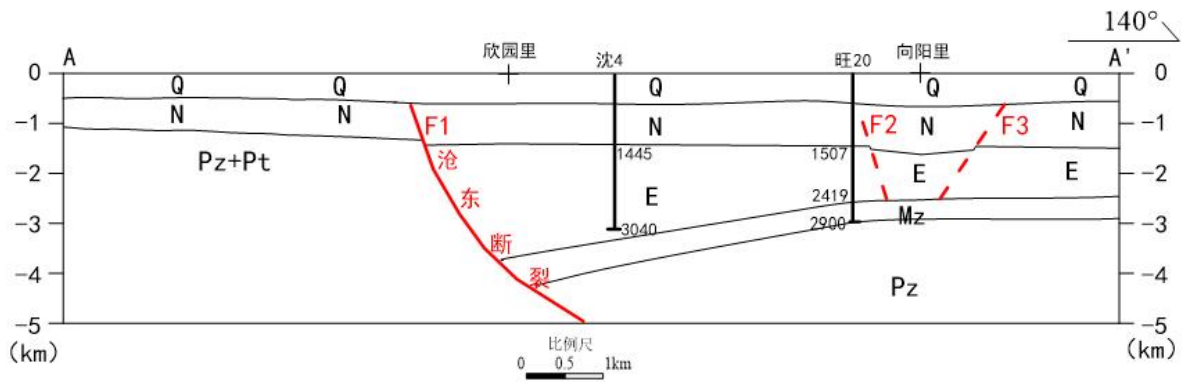

Fig. 9 AA 'Geological Profile

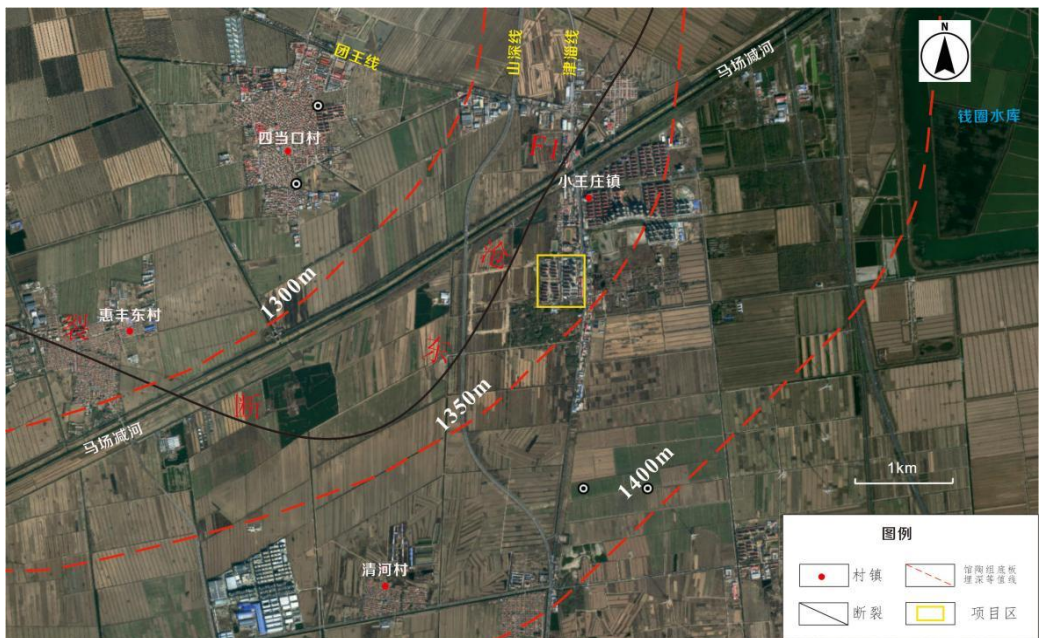

Fig. 10 Contour map of burial depth of the bottom plate of the Guantao Formation

## 4.3 Design and application of geothermal well

According to geophysical exploration and regional geological data, the geothermal well is designed. The geothermal well depth is 1450.00m (vertical depth) /1630.48m (oblique depth), and the production horizon is Neogene Guantao Formation. The mining depth is 1100-1400m (vertical depth). When the designed drawdown is ≤50m, the wellhead water temperature of geothermal well is ≥ 55 3℃, the wellhead water output is ≥ 80 5m/h, and the wellhead spacing between two wells is 5m. The plane position and geological design are shown in Fig. 11 and Fig.12.

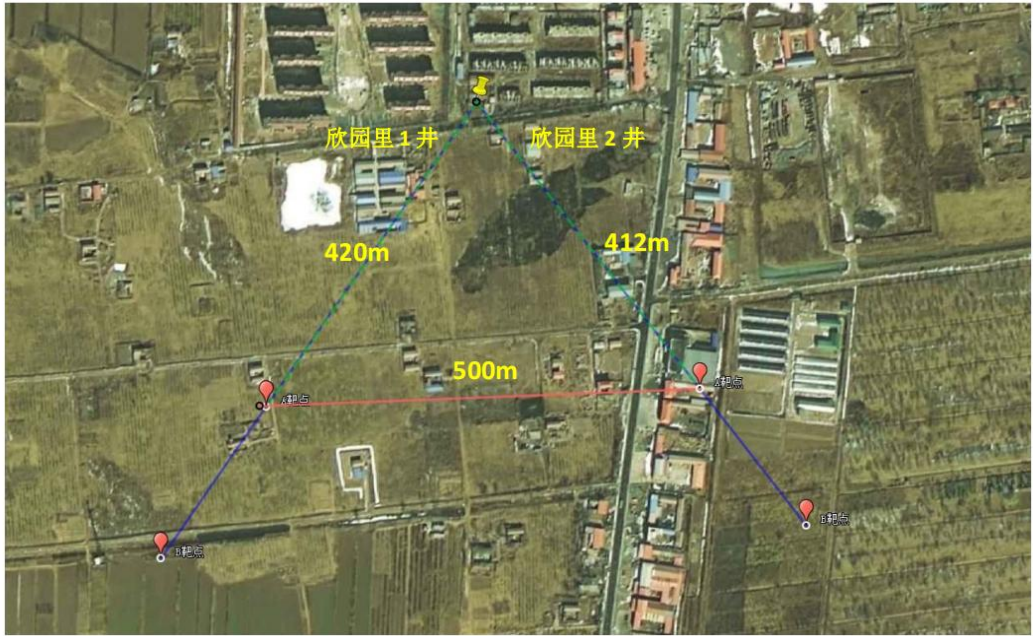

Fig.11 Deployment diagram of ZK1 well position in Banqiao depression

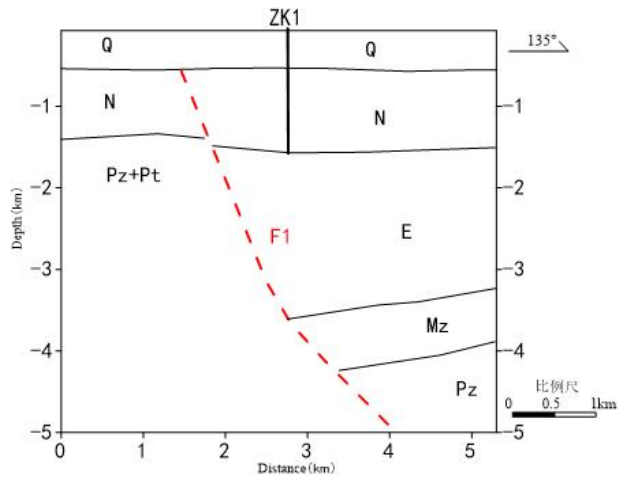

Fig. 12 Geological Design of ZK1 Well in Banqiao Depression

## 5 Conclusion

1. Based on the interpretation results of magnetotelluric sounding and the analysis of geological data, it is considered that the main developed fault in the working area is F1, namely Cangdong fault, which is located in the northwest of Xinyuanli. Cangdong fault is a deep fault in the region, with an overall dip angle of about 75, and the overall strike is NNE and southeast-east. It is a normal fault with a relatively rising west plate, and the buried depth of strata on both sides of the fault is very different.

2. The analysis of the interpretation results of magnetotelluric sounding shows that the following strata are developed in the working area from top to bottom: Quaternary, Neogene, Paleogene, Mesozoic, Paleozoic and Proterozoic. It is estimated that the buried depth of shallow strata in the working area of 5000m is as follows: the buried depth of Quaternary floor is 400~500m, that of Neogene floor is 1100 ~ 1600 m, that of Paleogene floor is 2400~3600m, that of Mesozoic floor is 2900~4200m, and that of Paleozoic floor is over 5000 m. However, according to the subsequent drilling verification, the buried depth of Quaternary floor in this area is 260-300m, that of Neogene Minghuazhen Formation is 1106-1294m, and that of Paleogene Guantao Formation is 1423-1540m, which is consistent with the geophysical interpretation results.

3. According to the results of geophysical exploration, the design and construction of geothermal wells are carried out. The drilling depth of geothermal wells is 1445.6m (vertical depth) /1630.0m (measured depth). When the depth is lowered by 36.8m during the productivity test, the water temperature is 51°C and the water volume is 89m$^3$/h, which shows a good geothermal resource potential.

## Declaration of Competing Interest

The authors report no declarations of interest.

## Acknowledgments

This paper is supported by the project Geophysical Exploration and Investigation of Southeast Block of xian county, Cangzhou City, Hebei Province by Sinopec Green Energy Geothermal Development Co., Ltd. Thanks to the reviewers for their valuable suggestions and guidance.

## References

Bostick, F. (1977). A simple almost exact method of magnetotelluric analysis. E., Ward, S.(ed.) in. Workshop on electrical methods in geothermal exploration: United States Geological Survey, Contract.

Bostick, F. ,1977. A simple almost exact method of magnetotelluric analysis. E., Ward, S.(ed.) in. Workshop on electrical methods in geothermal exploration: United States Geological Survey, Contract.

Constable, S. C., Parker, R. L., & Constable, C. G. ,1987. Occam's inversion: A practical algorithm for generating smooth models from electromagnetic sounding data. Geophysics, 52(3), 289-300.

Chave, A. D., Jones, A. G., Mackie, R., 2012. The Magnetotelluric Method (Theory and Practice) ‖ 3B. Description of the magnetospheric/ionospheric sources.

Lund, J. W., Freeston, D. H., Boyd, T. L., 2005. Direct application of geothermal energy: 2005 worldwide review. Geothermics, 34(6), 691-727.

Rosenkjaer, G. K., Gasperikova, E., Newman, G. A., 2015. Comparison of 3D MT inversions for geothermal exploration: Case studies for Krafla and Hengill geothermal systems in Iceland. Geothermics, 57(9), 258-274.

Shah, M., Sircar, A., Vaidya, D.,2015. OVERVIEW OF GEOTHERMAL SURFACE EXPLORATION METHODS. IJARIIE(4).

Swift, C. M., 1967. A magnetotelluric investigation of an electrical conductivity anomaly in the southwestern United States Massachusetts Institute of Technology.

Cai, J.T., Chen,X.B, Zhao, G.Z.,2010.Refined techniques for data processing and two-dimensional inversion in magnetotelluric I: Tensor decomposition and dimensionality analysis. Chinese J. Geophys. (in Chinese), 53(10): 2516-2526.(in Chinese).

Wang, Q.C et al.,2012 .Geophysical exploration of geothermal abnormal area in the Qiaogu-Kancaizhuang, Tianjin Binhai New Area." Geological Survey and Research 35(04), 304-309.(in Chinese).

Cao, X.G., Guo, Q.C.,Li LL.,2021. Application of MT method in the investigation and evaluation of geothermal resources on the east bank of the Yellow River in Yinchuan Plain. Western Exploration Project, 33(01), 141-144.(in Chinese).

Cao, X.L., Yan, L.J., Ch, Q.L, 2017. Application of Blind Source Separation Algorithm in Magnetotelluric Signal Denoising. Geophysical and Geochemical Computing Technology, 39(04), 456-464.(in Chinese).

Liu, Z.Y., Chen, X.B., Cai, J. T, et al.,2023. A magnetotelluric multi-scal and multi-period exploration method. Chinese J.Geophys,66(9),3761-3773.(in Chinese).

Chen, X.B., Zhao, G.Z., Tang, J, et al.,2005. An adaptive regularized inversion algorithmfor magnetotelluric data. Chinese J.Geophys,48(4) ,937-946.(in Chinese).

Deng, Y., XU,Y., FAN, Y.,e tal.,2024.Application of the magnetotelluric method in the Sichuan-Yunnan region—a review.Earth Science Frontiers,31(1):181-200.(in Chinese).

Han, G.M., Mou, L.G., Dong, Y.Q , et al.,2020.Cenozoic fault characteristics and petroleum geological significance in Banqiao slope area of Qikou Depression. Bulletin of Geological Science and Technology,39(6) :1-9.(in Chinese).

Li, B., Guo SW., Liu Y. ,2010. Forward modeling and damping least square inversion of magnetotelluric sounding based on MATLAB —— Taking layered one-dimensional media as an example. Inner Mongolia Petrochemical, 36(09), 35-36.(in Chinese).

Liu, J.X., Jin, Z J.,2004. Quantitative prediction of resource structure of Banqiao Beidagang reservoir-forming system. Journal of Daqing Petroleum Institute (01), 44-46+121.(in Chinese ).

Lei, Q., Ye G.F., Wu X.F, et al., 2024. Application of magnetotelluric sounding method in investigation and evaluation of deep carbonate thermal storage in Jizhong Depression. Geological Review, 70(02), 795-806.(in Chinese).

Li,S., Sun,X.L., Yang B.M,et al., 2023. Evaluation and analysis of thermal storage and recharge capacity of Neogene Guantao Formation sandstone in Tianjin. Geology of North China, 46(02), 38-44.(in Chinese ).

Li, S.T., Yue, D.D., Feng Z.L,et al., 2022. Sinoprobe and parameters study on deep karst geothermal reservoir in the Donglihu Area, Tianjin and its exploitable potential analysis. Geology in China, 49(6): 1732-1746.(in Chinese).

Ruan, C.X.,2018. Study on geothermal recharge of Wumishan Formation in Tianjin [D]Beijing: China Geo University (Beijing).(in Chinese).

Ruan, C.X., Feng, S.Y., David Shen, et al., 2017. Study on the recycling of geothermal resources in Tianjin Binhai New Area —— Research and demonstration of thermal storage and recharge technology in Guantao Formation. Journal of Geomechanics, 23(03), 498-506.(in Chinese ).

Ru, H.J., Liu D.L., Hu. H.C.,2018. Evaluation and Comprehensive Study of Geothermal Resources in Tianjin. Geological Survey of China, 5(02), 25-31.(in Chinese).

Sun, H.Q., Mao,X.,Wu, C.B.J,et al.,2024,Geothermal resources exploration and development technology:Current status and development directions.Earth Science Frontiers,31(1):400-411(in Chinese).

Sui, S.Q., Wang X.W., Zhou Z.Y, et al.,2019. Study on thermal storage characteristics of karst geothermal fields in Tianjin. Geology and Resources, 28(06), 590-594+569.(in Chinese).

Xu, S.Z., Liu B.,1995. Curve comparison method for magnetotelluric one-dimensional continuum inversion. Journal of Geophysics (05), 676-682.(in Chinese).

Shen, J., Hu, Y., 2005. Distribution and development prospect of geothermal resources in Tianjin coastal area.Exploration Science and Technology (06), 38-41+47.(in Chinese).

Song, F., Su N., Yao R.X,et al.,2016.Cenozoic Fault Structure and Hydrocarbon Accumulation Model in the Banqiao depression.Journal of South- west Petroleum University(Science&Technology Edition),38(2):49-58.(in Chinese).

Shi, Q.R., Fu, D.L., Wang, J.C., 2020. Geological Structure and Exploration Significance of Banqiao Mature Area in Qikou Depression. Urban Geology, 15(04), 432-437.(in Chinese ).

Tan J.,2015. Rhoplus Theory and AMT Dead Band Correction [D]Changsha:Central South University.

Tang, Y.X., Lin, J.W., Li.Q, et al., 2024. Occurrence regularity of deep geothermal resources in northern Tianjin Binhai geothermal field. Geology of North China, 47(01), 77-84.(in Chinese).

Tang, Y.X., Lin, J.W., Li.Q, et al., 2023. Exploration of carbonate thermal reservoir in Hangu fault zone and its significance for deep geothermal exploration. Geological Review, 69(S1), 95-96.(in Chinese).

Wang, S.J.，Shi,Y.Z.,Fang, C.H,et al.,2024. Status and development trends of geothermal development and utilization in oilfields of China.Lithologic Reservoirs,36(2):23-32.(in Chinese).

Wu, J.W., Hu,X.Y., Huang.G.S, et al.,2023. Present situation and prospect of electromagnetic exploration of geothermal resources. Acta Geologica Sinica, 44(01), 191-199.(in Chinese).

Xu, Y.C., Zhao,N., Qin, C, et al.,2015. One-dimensional Adaptive Regularization Inversion of Large Fixed Source Transient Electromagnetic. Geology and Exploration, 51(02), 360-365.(in Chinese ).

Yue, D.D., Jia, X.F., Zhang, Q.X, et al.,2023. Isotopic characteristics of thermal storage fluid of Wumishan Formation in Jixian system of Shanlingzi geothermal field in Tianjin and its indicative significance. Geology of North China, 46(02), 45-50.(in Chinese).

Yue, D.D.,Li, S.T., Jia, X.F., et al.,2020. Hydrochemical characteristics of thermal storage fluid of Wumishan Formation in Shanlingzi geothermal field, Tianjin. Science, Technology and Engineering, 20(36), 14847-14853.(in Chinese).

Yang, Y.H., Wu, X.M., Yang, L.F, et al. ,2020. Application of Bostick inversion in airborne electric field magnetotelluric method. Western Prospecting Project, 32(06), 137-139.(in Chinese).

Yin, Y .T., Wei, W.B., Ye, G.F, et al., 2012.An improved GB decomposition method based on geneticalgorithm. ChineseJ.Geophys, 55(2) :671-682 .(in Chinese).

Yao, Y.H., Jia,X.F., Li, S.T et al., 2022. Heat flow determination of Archean strata under the karst thermal reservoir of D01 well in Xiong'an New Area. Geology in China, 49(6): 1723-1731.(in Chinese).

Zhang Baiming.,2007. Research on Medium and Low Temperature Geothermal Thermal Storage Engineering [D]Beijing:China University of Geosciences (Beijing).(in Chinese).

Zhang, J.T., Zhou. J., Wang. X.B, et al.,2015. Searching for Lagrange Multiplier of One-dimensional Magnetotelluric Occam Inversion. Geophysical and Geochemical Computing Technology, 37(06), 687-692.(in Chinese).

Zhang, K., Dong, H., Yan, Y., et al.,2013. A parallel 3D inversion method of nonlinear conjugate gradient of magnetotelluric field. Journal of Geophysics (011), 056.(in Chinese).

Zhao, J.L., Chen., T.Z., Max, Z ., et al., 2010. Application of MT method in the investigation of geothermal resources in Kaifeng Depression. Geophysical and Geochemical Exploration, 34(02), 163-166.(in Chinese).